\documentclass[9pt,shortpaper,twoside,web,letterpaper]{IEEEtran} 
\usepackage{cite}
\usepackage{amsmath,amssymb,amsfonts,amsthm}
\usepackage{algorithm, algorithmic}
\usepackage{graphicx}
\usepackage{textcomp}
\def\BibTeX{{\rm B\kern-.05em{\sc i\kern-.025em b}\kern-.08em
    T\kern-.1667em\lower.7ex\hbox{E}\kern-.125emX}}
\usepackage{graphicx}
\usepackage{textcomp}

\usepackage{pgfplots}

\usepackage{float}
\usepackage{tikz}
\usetikzlibrary{arrows,%
                plotmarks, calc}
\usetikzlibrary[positioning, calc, intersections, backgrounds]

\usepackage[utf8]{inputenc}
  
\newcommand{\D}{\ensuremath{\mathcal D}}

\usepackage{xcolor}
\usepackage{xspace}
\usepackage{soul}
\usepackage{mathtools}
\usepackage{makecell}
\usepackage{cite}
\usepackage{nicefrac}
\usepackage{tcolorbox}
\usepackage{tcolorbox}

\newcommand{\ra}{\rightarrow}

\newcommand{\ie}{\unskip, i.\,e.,\xspace}
\newcommand{\eg}{\unskip, e.\,g.,\xspace}

\newcommand{\sut}{\text{s.\,t.\,}}

\newcommand{\wrt}{w.\,r.\,t.\xspace}

\newcommand{\N}{\ensuremath{\mathbb N }}

\newcommand{\R}{\ensuremath{\mathbb R }}

\newcommand{\X}{\ensuremath{\mathbb X }}
\newcommand{\Y}{\ensuremath{\mathbb Y }}
\newcommand{\F}{\ensuremath{\mathbb F }}
\newcommand{\U}{\ensuremath{\mathbb U }}

\newcommand{\sm}{\ensuremath{\setminus}}

\newcommand{\co}{\ensuremath{\overline{\text{co}}}}

\newcommand{\eps}{\ensuremath{\varepsilon}}

\newcommand{\ball}{\ensuremath{\mathcal B}}

\newcommand{\blue}[1]{\textcolor{blue}{#1}}

\definecolor{dgreen}{rgb}{0.0, 0.5, 0.0}

\newcommand{\spc}{\ensuremath{\,\,}}
\DeclareMathOperator*{\argmin}{arg\,min}

\newcommand{\norm}[1]{\left\lVert#1\right\rVert}  
\newcommand{\abs}[1]{\left\lvert#1\right\rvert}
\newcommand{\scal}[1]{\left\langle#1\right\rangle}
\newcommand{\set}[1]{\mathbb #1}

\makeatletter
\newcommand{\subalign}[1]{%
	\vcenter{%
		\Let@ \restore@math@cr \default@tag
		\baselineskip\fontdimen10 \scriptfont\tw@
		\advance\baselineskip\fontdimen12 \scriptfont\tw@
		\lineskip\thr@@\fontdimen8 \scriptfont\thr@@
		\lineskiplimit\lineskip
		\ialign{\hfil$\m@th\scriptstyle##$&$\m@th\scriptstyle{}##$\crcr
			#1\crcr
		}%
	}
}
\makeatother

\newtheorem{thm}{Theorem}
\newtheorem{dfn}{Definition}
\newtheorem{lem}{Lemma}
\newtheorem{exmp}{Example}
\newtheorem{asm}{Assumption}
\newtheorem{rem}{Remark}

\begin{document}
\title{On inf-convolution-based robust practical stabilization under computational uncertainty}
\author{Patrick Schmidt$^1$, Pavel Osinenko$^{1,2}$, Stefan Streif$^{1}$
\thanks{©2021 IEEE. Personal use of this material is permitted. Permission from IEEE must be obtained for all other uses, in any current or future media, including reprinting/republishing this material for advertising or promotional purposes, creating new collective works, for resale or redistribution to servers or lists, or reuse of any copyrighted component of this work in other works.}
\thanks{This work was partially funded by the European Union, European Social Fund ESF, Saxony.}
\thanks{$^1$Technische Universit\"at Chemnitz, Automatic Control and System Dynamics Lab, 09126 Chemnitz, Germany.} 
\thanks{$^2$Computational and Data Science and Engineering Center, Skolkovo Institute of Science and Technology, 143026 Moscow, Russia.}
\thanks{\textit{Corresponding author: Stefan Streif (stefan.streif@etit.tu-chemnitz.de).}} 
}

\maketitle

\begin{abstract}
	This work is concerned with practical stabilization of nonlinear systems by means of inf-convolution-based sample-and-hold control.
	It is a fairly general stabilization technique based on a generic non-smooth control Lyapunov function (CLF) and robust to actuator uncertainty, measurement noise, etc.
	The stabilization technique itself involves computation of descent directions of the CLF.
	It turns out that non-exact realization of this computation leads not just to a quantitative, but also qualitative obstruction in the sense that the result of the computation might fail to be a descent direction altogether and there is also no straightforward way to relate it to a descent direction.
	Disturbance, primarily measurement noise, complicate the described issue even more.
	This work suggests a modified inf-convolution-based control that is robust \wrt system and measurement noise, as well as computational uncertainty.
	The assumptions on the CLF are mild, as \eg any piece-wise smooth function, which often results from a numerical LF/CLF construction, satisfies them.
	A computational study with a three-wheel robot with dynamical steering and throttle under various tolerances \wrt computational uncertainty demonstrates the relevance of the addressed issue and the necessity of modifying the used stabilization technique.
	Similar analyses may be extended to other methods which involve optimization, such as Dini aiming or steepest descent.
\end{abstract}


\section{Introduction} \label{sec:intro}

Since not every nonlinear system can be asymptotically stabilized by a static continuous feedback \cite{Brockett1983-stabilization}, a great amount of research has been conducted in the search for alternative methods which include time-varying, dynamical and discontinuous control laws \cite{Astolfi1996-discontinuous}, \cite{Aastrom2013-adaptive}, \cite{Clarke2009-slid-mode-stab}, \cite{Filippov2013-discont-dyn-sys}, \cite{Fontes2003-opt-ctrl-discont}, \cite{Leith2000-survey}, \cite{Morin1999-design}.
In this work, we focus specifically on discontinuous control laws due to their relatively simple design (cf. sliding-mode control) as compared to the case of time-varying or dynamical controls whose design might be somewhat involved (compare \eg \cite{Bloch1996-stabilization} with \cite{Pomet1992-dyn-stabilization}).
Since a discontinuous control law leads, in general, to a closed-loop dynamical system with a discontinuous right-hand side, special attention must be paid to the treatment of system trajectories.
A good overview of generalized notions of the system trajectory in such cases was done by Cortes \cite{Cortes2008-discont-dyn-sys}.
One may implement the discontinuous control law in the sample-and-hold (SH) manner, in which the control actions are held constant during predefined time samples.
This enables ``standard'' Carathéodory system trajectories at the cost of given up asymptotic stability for practical stability which describes convergence to any predefined vicinity of the equilibrium within finite time \cite{Clarke2004-lyapunov}. 
Practical stability, although being a weaker form of stability than the asymptotic one, is still widely applicable.

This work addresses practical stabilization with the use of a control Lyapunov function (CLF).
The latter can be obtained by various techniques \cite{Baier2012-linear}, \cite{Baier2014-num-CLF}, \cite{Bianchini2018-merging}, \cite{Giesl2015-review}, \cite{Malisoff2009-constructing-strict-LF}.
The resulting CLF is often nonsmooth (in general, this is the case when the system fails to satisfy Brockett's condition) \cite{Brockett1983-stabilization}.
This property differentiates the current work from other existing ones, such as \cite{delaPena2008-lyapunov}, where local differentiability is assumed.
Stabilizing control actions can be determined from the CLF in different ways \cite{Braun2017-SH-stabilization-Dini-aim} \eg steepest descent, infimum convolution (InfC),  Dini aiming \cite{Kellett2004-Dini-aim}, \cite{Kellett2000-Dini-aim} and optimization-based feedback.
Robustness properties of some of these SH stabilizing controls were extensively studied \cite{Clarke2011-discont-stabilization}, \cite{Clarke1997-stabilization}, \cite{Sontag1999-stability-disturb}.
It is mainly the measurement noise that might complicate the stabilization due to the phenomenon called ``chattering'' \cite{Sontag1999-stability-disturb} whereas the model and actuator uncertainty can be addressed straightforwardly. 
The issue may be tackled by various means, such as \eg the so called ``internal tracking controller'' \cite{Ledyaev1997-stabilization-meas-err}.
On the other hand, the InfC control possesses a natural robustness with regards to the measurement noise \cite{Sontag1999-stability-disturb}.
In this work, we focus specifically on this kind of control.
The main challenge is that the optimization problems, which are involved in the computation of the InfC stabilizing control actions, \textit{cannot in general be solved exactly}.
This non-exactness can be understood as a \textit{computational uncertainty}.
The importance of addressing it was stated in several works \eg \cite[Problem~8.4]{blondel2009-unsolved}, \cite{Gao2014-descriptive}.

This works starts with a nominal system under the InfC feedback $\kappa$ in the SH mode.
The transition from a system $\dot x = f(x, \kappa(x))$ to the one $\dot x = f(x, \tilde \kappa(x))$, where $\tilde \kappa$ denotes the InfC feedback in the SH mode under non-exact computation, was addressed in \cite{Osinenko2018-practical-SH}.
\textit{The goal of this work is to fuse the result of \cite{Osinenko2018-practical-SH} with robustness \wrt measurement noise and system disturbance, which is a challenging task.} 
Furthermore, the aim of the paper is a verified analysis of nonlinear systems extended by a measurement error and system disturbance.
Verified here means that an algorithm is derived which enables computing necessary bounds on the sampling time, at least in principle.
The central result, namely, a theorem on robust practical stabilization by InfC under computational uncertainty is presented in Section \ref{sec:prac-stab}, followed by a case study in Section \ref{sec:case-study}.

The core text will list technical lemmas and the main theorem with its proof sketch, while the detailed proofs are provided in the appendix.

\textit{Notation}: $\ball_R(x)$ describes a ball with radius $R$ at $x$ \ie $\ball_R(x) := \{ x: \norm x \leq R \}$ and $\ball_R$ means that $x = 0$; $\co (\set X)$ denotes the closure of the convex hull of a set $\set X$; $\norm \bullet$ denotes the Euclidean norm; $\R_{> 0}, \R_{\ge 0}$ are the sets of positive, respectively, non-negative real numbers.

\section{Preliminaries} \label{sec:prelim}

\subsection{System description and assumptions}

This work addresses practical stabilization of an uncertain nonlinear system in the following form:
\begin{equation}
	\label{eqn:sys}
	\dot x  = f(x, \kappa(\hat x)) + q,	
\end{equation}
where $x, \hat x \in \R^n$ denote the state and, respectively, its measurement, $q: \R_{\ge 0} \ra \R^n$ is a (time-varying) disturbance, $\kappa: \R^n \ra \R^m$ is a control law that only has access to the measured state $\hat x$.
We assume that the admissible control actions are in some compact input constraint set $\U$.

The following is assumed about \eqref{eqn:sys}.
\begin{asm}[System properties]
	\label{asm:sys-props}
	\hspace{1pt}
	\begin{itemize}
		\item (disturbance boundedness) there exist numbers $\bar e, \bar q$ \sut $\forall t \ge 0$ $\norm{x(t) - \hat x(t)} \le \bar e$ and $\norm{q(t)} \le \bar q$;
		\item (Lipschitz property) for any $z \in \R^n$ and $\omega > 0$ there exists $L_f = L_f(z, \omega) > 0$ such that for all $x, y \in \ball_\omega(z)$ and for all $u \in \U$,
\begin{equation} 
	\label{eqn:Lip-cond}
	\norm{ f(x, u) - f(y, u) } \le L_f \norm{x-y}.
\end{equation}
	\end{itemize}
\end{asm}

Notice that a system with a bounded actuator uncertainty $p(t), p: \R_{\ge 0} \rightarrow \R^m$ of the form
\begin{equation}
	\dot x = f(x, \kappa(\hat x) + p(t)) 
\end{equation}
can be transformed into the form \eqref{eqn:sys}, using \eqref{eqn:Lip-cond}, and so we omit actuator uncertainty from now on.

\subsection{Controller description}

Firstly, as discussed in the introduction, we implement the control law $\kappa$ in the SH mode as follows:
\begin{equation} 
	\label{eqn:sys-SH}
	\begin{split}
		& \dot x = f(x, u_k) + q, \\
		& t \in [ k \delta, (k + 1) \delta], u_k \equiv \kappa(\hat x (k \delta)), k \in \N,
	\end{split}
\end{equation}
where $\delta$ is the sampling time (for simplicity of further derivations assumed constant).
The starting point of practical stabilization is a proper, positive-definite, locally Lipschitz continuous control Lyapunov function (CLF) $V: \R^n \ra \R$ that satisfies the following condition \cite{Clarke1997-stabilization}: for each compact set $\X \subseteq \R^n$, there exists a compact $\U(\X) \subseteq \U$ such that
\begin{equation} 
	\label{eqn:decay}
	\forall x \in \X \inf_{ \theta \in \co (f(x, \U(\X))) } \D_\theta V(x) \le -w(x),
\end{equation}
where $w: \R^n \ra \R$ is a continuous non-negative function with $\spc x \ne 0 \implies w(x) > 0$.
In \eqref{eqn:decay}, $\D_\theta V(x)$ denotes the \emph{generalized directional lower derivative} in a direction $\theta \in \R^n$, defined by
\begin{equation} 
	\label{eqn:dini-der}
	\D_\theta V(x) \triangleq \liminf_{ \mu \ra 0^+ } \frac{ V(x + \mu \theta) - V(x) }{\mu}.
\end{equation}
Practical stabilization is defined in the following way:
\begin{dfn}[Practical stabilization]
	\label{dfn:practical-stabilization}
	Consider a system \eqref{eqn:sys-SH} with $e \equiv 0$ and $q \equiv 0$.
	Then, a control law $u = \kappa(x)$ practically stabilizes \eqref{eqn:sys-SH} in the sample-and-hold mode, if for all $r,R$ with $R > r > 0$, there exists a sufficiently small sampling time $\delta > 0$ such that any closed-loop trajectory $x(t)$ with $x(0) \in \ball_R$, is bounded and enters and remains in $\ball_r$ after a time $T$ depending uniformly on $r$ and $R$.
\end{dfn}
To practically stabilize the system \eqref{eqn:sys-SH}, the control action $u_k$ is computed at each time step $k \in \N$.
There are different techniques for this task as discussed in the introduction, and we focus on InfC.
First, consider the following \emph{inf-convolution} \cite{Clarke2008-nonsmooth-analys} of $V$:
\begin{equation} 
	\label{eqn:InfC}
	V_\alpha(x) := \inf_{ y \in \R^n } \left\{ V(y) + \frac{1}{2 \alpha^2} \norm{y - x}^2 \right\}, \spc \alpha \in (0,1).
\end{equation}
The above equation is also known as Moreau-Yosida regularization \cite{Lemarechal1997-practical}.
For a $y_\alpha(x)$, a corresponding minimizer for \eqref{eqn:InfC}, the vector
\begin{equation} 
	\label{eqn:subgradient-zeta}
	\zeta_\alpha(x) := \frac{ x - y_\alpha(x) }{\alpha^2}
\end{equation}
happens to be a \textit{proximal subgradient} of $V$ at $x$ in the sense that
\begin{equation} 
	\label{eqn:subgradient-property}
	V(z) \ge V( y_\alpha(x) ) + \scal{ \zeta_\alpha(x), z - y_\alpha(x) } - \frac{\norm{ z - y_\alpha(x) }^2}{2 \alpha^2} 
\end{equation}
holds for all $z \in \R^n$.

The core of the InfC control under exact optimization is the following property:
\begin{equation} 
	\label{eqn:subgrad-property-scal}
	\scal{ \zeta, \theta } \le \D_\theta V(x),
\end{equation}
which holds for all proximal subgradients $\zeta$ of $V$ at each point $x$ and for any direction $\theta$.
The corresponding control algorithm can be found \eg in \cite{Clarke1997-stabilization}.
Namely, at each time step $t_k = \delta k$, compute $y_\alpha (x_k)$ and $\zeta_\alpha (x_k)$ based on the current state $x_k$.
Then, determine the control action $u_k$ by 
\begin{equation}
	u_k \in \set U_k, \set U_k := \argmin_{u \in \U} \scal{\zeta_\alpha (x_k), f(x_k, u)}.
\end{equation}
Now, under \textit{computational uncertainty}, the minimizer $y_\alpha(x)$ has to be substituted with an approximate minimizer $y_\alpha^\eps(x)$, which, for some optimization accuracy $\eps_x > 0$ (that may depend on $x$), yields:
\begin{equation} 
	\label{eqn:InfC-approx-optim}
		\forall x \in \R^n: \spc V(y_\alpha^\eps(x)) + \frac{1}{2 \alpha^2} \norm{ y_\alpha^\eps(x) - x }^2 \le V_\alpha(x) + \eps_x.
\end{equation}
The control action $\kappa_x^\eta$ also yields merely an approximate condition of the form
\begin{equation} 
	\label{eqn:ctrl-approx-optim}
	\scal{ \zeta_\alpha^\eps(x), f(y_\alpha^\eps(x), \kappa_x^\eta) } \le \inf_{ u \in \U(\Y) } \scal{ \zeta_\alpha^\eps(x), f(y_\alpha^\eps(x), u) } + \eta_x,
\end{equation}
where $\eta_x > 0$ denotes the respective optimization accuracy and $\U(\Y) \subseteq \U$ is the set of admissible control actions for a given compact set $\Y$ containing $y_{\alpha}^\eps (x)$, so that \eqref{eqn:decay} holds for all $y \in \Y$.
Notice that the vector
\begin{equation} 
	\label{eqn:subgrad-zeta-approx-optim}
	\zeta_\alpha^\eps(x):=\frac{ x - y_\alpha^\eps(x) }{\alpha^2}
\end{equation}
is \textit{not}, in general, a proximal subgradient.
Consequently, the property \eqref{eqn:subgrad-property-scal}, which is absolutely crucial in InfC, \textit{cannot be used directly under computational uncertainty}.

In this work, we are concerned with \textit{computational uncertainty} and do not assume exact knowledge of $y_\alpha(x)$ for given $\alpha$ and $x$.
Instead, we use \textit{approximate minimizers} in the sense of the following:
\begin{lem} 
	\label{lem:1-eps-minimizer-vicinity}
	Let $R > 0$, $\alpha \in (0,1)$ and $\eps > 0$.
	Then, for all $x \in \ball_R$ there exists an $\eps$-minimizer $y_\alpha^\eps(x)$ for \eqref{eqn:InfC} satisfying 
	\begin{equation} 
		\label{eqn:vicinity-y-x}
		\norm{ y_\alpha^\eps(x) - x } \le (2 \bar V)^{\nicefrac{1}{2}} \alpha,
	\end{equation}		
	where $\bar V:=\sup_{ \norm x \le R } V(x)$.
\end{lem}

The inf-convolution has the following approximation property under approximate minimizers:
\begin{lem} 
	\label{lem:2-accuracy-InfC}
	Under the conditions of Lemma \ref{lem:1-eps-minimizer-vicinity}, for any $\eps_1 > 0$, an $\eps > 0$ and an $\alpha \in (0,1)$ can be chosen for $y_\alpha^\eps(x)$ so as to satisfy, for all $x \in \ball_R$, the following property:
	\begin{equation} 
		\label{eqn:V-V-alpha-approx}
		V_\alpha(x) \le V(x) \le V_\alpha(x) + \eps_1.
	\end{equation}	
\end{lem}

In the following, we refer to the control law, whose control actions are determined via \eqref{eqn:InfC-approx-optim} and \eqref{eqn:ctrl-approx-optim} as \textit{uInfC}, a shorthand for \emph{InfC control under computational uncertainty}.
We subsequently pursue \textit{robust practical stabilization under computational uncertainty} in the following sense (cf. \cite{Ledyaev1997-stabilization-meas-err}):
\begin{dfn}[Semiglobal robust practical stabilization by uInfC] 
	\label{def:robust-pract-stab}
	An \textit{uInfC} is said to robustly practically stabilize \eqref{eqn:sys} in the SH mode \eqref{eqn:sys-SH} if, for each $R$ and $r \in (0, R)$, there exist numbers
	\begin{equation*}
		\begin{array}{lll}
			\tilde e = \tilde e(r, R) > 0, & \tilde q = \tilde q (r, R) > 0, & \\
			\tilde \eta = \tilde \eta(r, R, x) > 0, & \tilde \eps = \tilde \eps (r, R, x) > 0, & \tilde \delta = \tilde \delta (r, R) > 0,
		\end{array}
	\end{equation*} 
	depending uniformly on $r,R$ and $x \in \R^n$, such that if the following properties hold:
	\begin{itemize}
		\item the sampling time satisfies $\delta \le \tilde \delta$;
		\item the accuracies in \eqref{eqn:InfC-approx-optim} and \eqref{eqn:ctrl-approx-optim} are bounded as $\eps_{\hat x_k} \leq \tilde \eps$, $\eta_{\hat x_k} \leq \tilde \eta$, where $\hat x_k$ is the sampled measured state at a step $k \in \N$;
		\item the bounds on the measurement error and disturbance satisfy $\bar e \leq \tilde e$ and $\bar q \leq \tilde q$, respectively;
	\end{itemize}
	then, any closed-loop trajectory $x(t), \spc t \ge 0$, $x(0) = x_0 \in \ball_R$ is bounded and there exists $T$ \sut $x(t) \in \ball_r, \spc \forall t \ge T$.
\end{dfn}

\begin{rem} 
	\label{rem:robust-pract-stab-unif}
	The considered optimization accuracy bounds $\tilde \eps, \tilde \eta$ in Definition \ref{def:robust-pract-stab} depend on the current sampled measured state $\hat x_k$ at a sample step $k \in \N$.
	The derived results of this work allow also a uniform choice of $\tilde \eps, \tilde \eta$ \ie independent of the current sampled measured state (see Remark \ref{rem:eps-state-indep}).
\end{rem}

The next section presents the main theorem on practical robust stabilization under computational uncertainty.

\section{Robust practical stabilization under computational uncertainty} \label{sec:prac-stab}

The work \cite{Osinenko2018-practical-SH} showed practical stabilization by InfC using a certain additional assumption on the given CLF.
Here, we relax this assumption to the following version:
\begin{asm} 
	\label{asm:separable-edges}
	For all compact sets $\Y, \F \subset \R^n$ and for all $\nu, \chi > 0$ there exist $\tilde \Y \subseteq \Y, \mu \ge 0$ such that:
	\begin{enumerate} 
		\item 
		for each $\tilde y \in  \tilde \Y, \theta \in \F$ and $\forall \mu' \in (0, \mu]$ it holds that
		\begin{equation} 
			\label{eqn:loc-hom-cond}
			\abs{ \frac{ V(\tilde y + \mu' \theta) - V(\tilde y) }{\mu'} - \D_\theta V(\tilde y) } \le \nu;
		\end{equation}
		\item for each $y \in \Y$ there exists $\tilde y \in \tilde  \Y$ such that
		\begin{equation} 
			\label{eqn:neighb-pt-cond}
			\norm{y - \tilde y} \le \chi.
		\end{equation}
	\end{enumerate}
\end{asm}

\begin{rem}
	The first part in Assumption \ref{asm:separable-edges} contains a local homogeneity condition for all points $\tilde y \in \tilde{\Y}$ \ie $V$ is globally lower Dini differentiable and the $\liminf$ in \eqref{eqn:dini-der} is locally uniform, as stated in \cite{Osinenko2018-practical-SH}.
The second part in Assumption \ref{asm:separable-edges} covers all points in $\Y$, which do not satisfy \eqref{eqn:loc-hom-cond}.
	On the contrary, Assumption 1  in \cite{Osinenko2018-practical-SH} contains only part 1 of Assumption \ref{asm:separable-edges}.
Nevertheless, stabilization is also possible, if \eqref{eqn:loc-hom-cond} does not hold for all $y \in \Y$ but rather $\tilde{\Y} \subset \set A \subset \R^n$, where the complement of $\set A$, denoted by $\set A_0 := \R^n \sm \set A$, is given as a set with measure zero and $\tilde{\Y} \not\subset \ball_{\chi}(\set A_0) := \{ y \in \R^n: \norm{y - \set A_0} \leq \chi \}$. 
In Assumption \ref{asm:separable-edges}, $\Y \sm \tilde{\Y}$ is such a set of measure zero, and part 2 secures a global stabilization result.
If $y_\alpha^{\eps} (\hat x)$ lies in such a set, Assumption 1 in \cite{Osinenko2018-practical-SH} would not be satisfied.
	It can be shown that, for instance, any piece-wise affine function satisfies this assumption (a small demonstrative example is given in the appendix).
Such CLFs arise \eg in triangulation-based numerical constructions of Lyapunov functions \cite{Baier2014-num-CLF}.
Therefore, the above assumption is fulfilled by a larger set of CLFs, than Assumption 1 in \cite{Osinenko2018-practical-SH}, namely by all CLFs with countable number of sets of zero measure.
	Assumption \ref{asm:separable-edges} is interpreted algorithmically in the sense that we can always be provided with a point $\tilde y$ for a $\chi$ that is specified later.
\end{rem}

We can now state the main result.
\begin{thm} 
	\label{thm:pract-stab}
	Consider the system \eqref{eqn:sys} and let Assumption \ref{asm:sys-props} hold.
	Let $V$ be a CLF satisfying \eqref{eqn:decay} and Assumption \ref{asm:separable-edges}. 
	Then, \eqref{eqn:sys} can be practically robustly stabilized by uInfC control in the SH mode \eqref{eqn:sys-SH} in the sense of Definition \ref{def:robust-pract-stab}.
\end{thm}

\begin{rem}
	Theorem \ref{thm:pract-stab} ensures robust practical stability of \eqref{eqn:sys-SH} up to prescribed precision in terms of the parameters $R$ and $r$, if the bounds on sampling time, system disturbance, measurement error and optimization accuracy are fulfilled.
	Since the proof is constructive, the derived bounds on the sampling time can be computed, at least in principle, though might be conservative depending on the system, given CLF and decay rate.	
	Nevertheless, they can be adapted to obtain more suitable bounds.
	Some ideas are discussed in Section \ref{sec:case-study}.
\end{rem}

Now, a sketch of the proof is presented.
The whole proof can be found in the appendix. 
It is also the basis for the presented algorithm.
\begin{proof}\textit{(Sketch)} 
	The first part of the proof is concerned with deriving some a priori bounds based on the given starting and target ball radii, say, $R$ and $r$.
	Among these bounds, is the one on the trajectory overshoot and, most importantly, the one on the guaranteed decay rate of $V_\alpha$ until the state reaches the target ball.
	As one can see, in InfC, we work effectively with the inf-convolution $V_\alpha$ instead of the original CLF $V$.
	
	In the second part, to actually show sample-to-sample decay of $V_\alpha$, we need to derive particular bounds on the optimization accuracies $\eta_{\hat x_k}$ and $\eps_{\hat x_k}$ with special care.
	This process is complicated by the fact that we do not have access to the true state, but to only an estimate thereof, the $\hat x$.

	In the third part, we use a property of $V_\alpha$ analogous to Taylor series expansion in smooth analysis (keep in mind, we work with non-smooth tools all along).
	Expressing some bounds on the inter-sample system trajectory, we can show that $V_\alpha$ decays sample-to-sample to a limit that guarantees that the true state $x$ enters and never leaves the target ball $\ball_r$ provided that some additional conditions on the sampling time and optimization accuracies hold.
	This part is somewhat tedious, but made possible by exploiting Assumption \ref{asm:separable-edges}.	
\end{proof}

Algorithm \ref{alg:decay} summarizes the uInfC control procedure.
\begin{algorithm}[H]
	\caption{uInfC}
	\label{alg:decay}
	\begin{algorithmic}[1]
		\renewcommand{\algorithmicrequire}{\textbf{Input:}}
		\renewcommand{\algorithmicensure}{\textbf{Set:}}
		\REQUIRE System $\dot x = f(x, u) + q, \hat x = x + e$ and a CLF $V(x)$
		\ENSURE  Sampling time $\delta$
		\\ At $t_k = \delta k$:
			\STATE Measure $\hat x_k$
			\STATE Compute $y_\alpha^{\eps_x}(\hat x_k)$ via InfC \eqref{eqn:InfC-approx-optim} with accuracy at least $\eps_{\hat x_k}$
			\STATE Compute control action $\tilde \kappa^{\eta_x}_{\hat x_k}$ by \eqref{eqn:ctrl-approx-optim} with accuracy at least $\eta_{\hat x_k}$ using $\tilde y_\alpha^{\eps_x}(\hat x_k)$ from Assumption \ref{asm:separable-edges}
			\STATE Apply $\tilde \kappa^{\eta_x}_{\hat x_k}$ to the system and hold constant until the next sample $k+1$
	\end{algorithmic}
\end{algorithm}

In the following section, we study robust practical stabilization by uInfC of the so-called extended nonholonomic dynamic integrator (ENDI) which is essentially a model of a three-wheel robot with dynamical steering and throttle.
Such a model is a prototype of many real-world machines.

\section{Case study: Extended nonholonomic integrator} \label{sec:case-study}

A three-wheel robot with dynamical actuators of the driving and steering torques is described as follows \cite{Abbasi2017-backstepping,Pascoal2002-practical,Sankaranarayanan2009-switched}:
\begin{equation} 
	\label{eqn:ENDI}
	\tag {ENDI}
	\begin{split}
		\dot \varphi_1 &= \eta_1 \\
		\dot \varphi_2 &= \eta_2 \\
		\dot \varphi_3 &= \varphi_1 \eta_2 - \eta_1 \varphi_2 \\
		\dot \eta_1 &= u_1 \\
		\dot \eta_2 &= u_2.
	\end{split}
\end{equation}
The ENDI is essentially the Brockett's nonholonomic integrator 
\begin{equation} 
	\label{eqn:NI}
	\tag {NI}
	\dot \varphi = \underbrace{\begin{pmatrix}
		1 \\ 0 \\ -\varphi_2
	\end{pmatrix}}_{=: g_1(\varphi)} \omega_1 + \underbrace{\begin{pmatrix}
		0 \\ 1 \\ \varphi_1
	\end{pmatrix}}_{=: g_2(\varphi)} \omega_2
\end{equation}
with additional integrators before the control inputs.
A (locally semiconcave) CLF for \eqref{eqn:ENDI} can be computed via non-smooth backstepping as per \cite{Matsumoto2015-position}.
Namely, we set the state vector as $x = \begin{pmatrix} \varphi^\top & \eta^\top \end{pmatrix}^\top$ and
\begin{equation} 
	\label{eqn:ENDI-V}
	V(x) = \min_{\theta \in [0,2 \pi)} \left\{  \tilde F(\varphi; \theta) + \frac{1}{2} \norm{ \eta - \kappa(\varphi; \theta) }^2 \right\},
\end{equation}
where
\begin{equation} 
	\label{eqn:CLF-NI}
	\tilde F(\varphi; \theta) = \varphi_1^2 + \varphi_2^2 + 2 \varphi_3^2 - 2 \varphi_3 (\varphi_1 \cos \theta + \varphi_2 \sin \theta),
\end{equation}
and
\begin{equation} 
	\label{eqn:feedback-NI}
	\kappa(\varphi; \theta) = -\begin{pmatrix}
		\scal{\zeta(\varphi; \theta), g_1(\varphi)} \\ \scal{\zeta(\varphi; \theta), g_2(\varphi)}
	\end{pmatrix}, \zeta(\varphi; \theta) = \nabla_\varphi \tilde F(\varphi; \theta).
\end{equation}
Note that for the minimizer $\theta^\star$ of \eqref{eqn:CLF-NI}, $\tilde F(\varphi, \theta)$ reduces to the CLF given in \cite{Braun2017-SH-stabilization-Dini-aim} as
\begin{equation} 
	\label{eqn:CLF-NI-V}
	\tilde F(\varphi; \theta^\star) = \tilde V(\varphi) = \varphi_1^2 + \varphi_2^2 + 2 \varphi_3^2 - 2 \abs{\varphi_3} \sqrt{ \varphi_1^2 + \varphi_2^2 }.
\end{equation}

The results of simulation under different accuracies and disturbance bounds are presented in the following.
The initial condition is set to $x_0 = \begin{pmatrix} -1 & 0.5 & 0.2 & 0.1 & 0.1 \end{pmatrix}^\top$ and the set of admissible controls is given as $\U = [-3,3]$.
Furthermore, we set $\alpha = 0.1$ and $\delta = 10^{-4}$. \linebreak

\begin{figure}
\centering
  \includegraphics[width = 0.475\textwidth, trim = {0.7cm 0.7cm 0.7cm 0.7cm},clip]{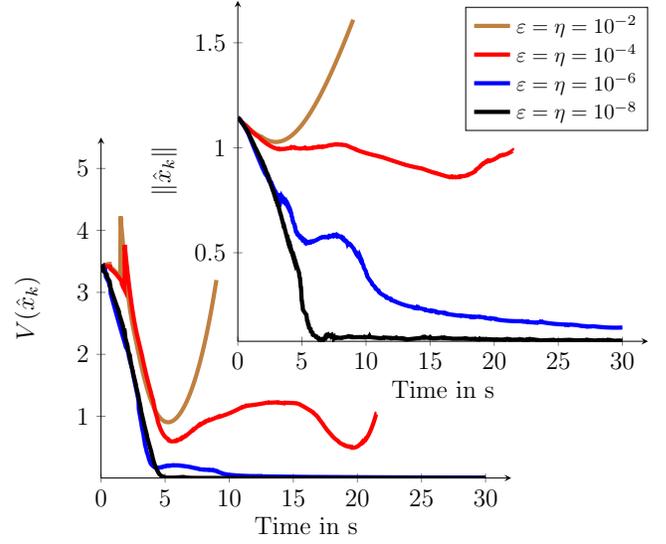}
	\caption{Norm of the state $\norm{\hat x_k}$ and Lyapunov function $V(\hat x_k)$ for different optimization accuracies.}
	\label{fig:gfx_epsilon}
\end{figure}

In the first simulation, the influence of the optimization accuracy on the state convergence is studied.
Fig. \ref{fig:gfx_epsilon} shows the CLF behavior and the norm of the states along with the controls for different values of $\eps_{\hat x}$ and $\eta_{\hat x}$, namely $\eps_{\hat x} = \eta_{\hat x} \in \{ 10^{-2}, 10^{-4}, 10^{-6}, 10^{-8} \}$, and $\bar q = \bar e = 0.5 \cdot 10^{-3}$.
It can be observed that insufficient accuracy ($\eps_{\hat x} = \eta_{\hat x} = 10^{-2}$ or $\eps_{\hat x} = \eta_{\hat x} = 10^{-4}$) leads to the loss of practical stability.
Higher accuracies lead to ever smaller vicinities of the origin that the state converges into.
This clearly demonstrates that computational uncertainty must be taken into account in practical stabilization.

\begin{figure}
\centering
  \includegraphics[width = 0.475\textwidth, trim = {0.7cm 0.7cm 0.7cm 0.7cm},clip]{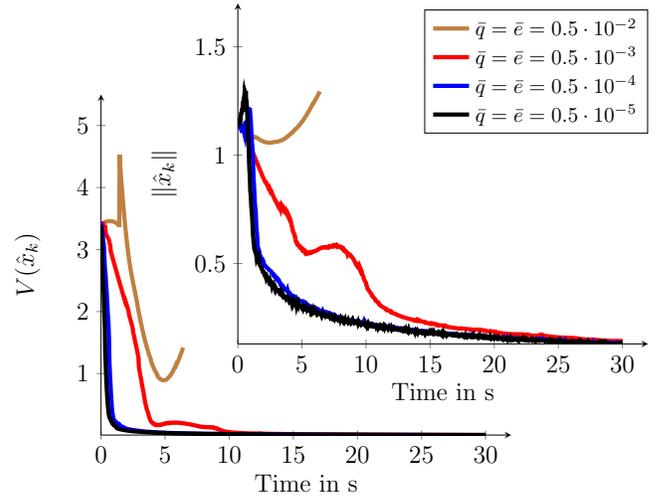}
	\caption{Norm of the state $\norm{\hat x_k}$ and Lyapunov function $V(\hat x_k)$ for different error and disturbance bounds.}
	\label{fig:gfx_error}
\end{figure}

In the second simulation, the influence of $\bar e$ and $\bar q$ is investigated.
We set $\bar e = \bar q \in \{ 0.5 \cdot 10^{-2}, 0.5 \cdot 10^{-3}, 0.5 \cdot 10^{-4}, 0.5 \cdot 10^{-5} \}$, and $\eps_{\hat x} = \eta_{\hat x} = 10^{-6}$. 
From Fig. \ref{fig:gfx_error} it can be observed, that the trajectory converges faster to the origin for smaller measurement errors and disturbance bounds.
For $\bar e = \bar q = 0.5 \cdot 10^{-2}$, the algorithm fails to stabilize the system.

Finally, it can be observed that the results only have small improvements for much higher restrictions on optimization accuracy and error bounds.
Based on the algorithm derived from the proof of Theorem \ref{thm:pract-stab} \ie Algorithm \ref{alg:computation-delta}, an upper bound for the sampling time can be stated as $\bar \delta = 0.23 \cdot 10^{-6}$ and for the optimization accuracy as $\eps_{\hat x} = 0.18 \cdot 10^{-6}$.
Thus, the computation of a verified bound on the sampling time is plausible, but rather conservative (which is somewhat expected). 
The computed bounds might be relaxed provided with some physical insight into the given system, such as maximum velocity of the respective differential equation, for instance. 
A more detailed discussion on this requires future work and goes beyond the
scope of the current one.

\begin{algorithm}[H]
	\caption{Upper bounds for sampling time, optimization accuracies and error bounds based on the proof of Theorem \ref{thm:pract-stab}}
	\label{alg:computation-delta}
	\begin{algorithmic}[1]
		\renewcommand{\algorithmicrequire}{\textbf{Input:}}
		\renewcommand{\algorithmicensure}{\textbf{Set:}}
		\REQUIRE System $\dot x = f(x, u) + q, \hat x = x + e$ and CLF $V(x)$
		\ENSURE  $R$, $r$, $\bar e$, $\bar q$, $\set U$
			\STATE Compute $\alpha_1(x)$, $\alpha_2(x)$ and $w(x)$ such that $\alpha_1(x) \leq V(x) \leq \alpha_2(x)$ and \eqref{eqn:w(x)} hold.
			\STATE Define $\varrho_V(x) = \alpha_1(x)$ and $\lambda_V(x) = \alpha_2^{-1}(x)$.
			\STATE Compute $\hat V$, $\hat R^*$, $\hat v$, $\hat r^*$, $\bar f$, $L_V$, $L_f$, and $\bar w$ according to part 1 in the proof of Theorem \ref{thm:pract-stab}.
			\item Compute upper bounds for $\eps_1$, $\alpha$, $\eta_x$, $\delta$, $\eps_{\hat x}$, $\chi$, $\bar e$, $T_\alpha$ based on \eqref{eqn:Bounds-1}-\eqref{eqn:Bounds-4} and \eqref{eqn:bound-5}.	
	\end{algorithmic}
\end{algorithm}

\section{Conclusion} \label{sec:concl}

This work was concerned with practical robust stabilization of nonlinear systems under computational uncertainty related to non-exact optimization. 
We showed that, under a mild assumption on the CLF, the InfC controller can robustly practically stabilize the given system even if the computations involved are merely approximate.
The result should be seen as complementary to the existing ones which are only concerned with robustness regarding system and measurement noise.
Summarizing, in addressing practical stabilization, computational uncertainty should be considered along with other uncertainties, especially in the cases where safety is crucial.

\section{Appendix}

\subsection{Demonstration of Assumption \ref{asm:separable-edges}}

\begin{exmp} 
	\label{exm:abs-fnc}
	Consider $V(x) = \abs x$.
	Let $\Y, \F$ be given and choose $\tilde{\Y} = \Y \sm \left[ -\nicefrac \chi 2, \nicefrac \chi 2 \right]$.	
	Without loss of generality, $y > 0$ is considered (the other cases are treated analogously).
	Since $\F$ is compact, there exist bounds such that for all $\theta \in \F: \theta_{\text{min}} \leq \theta \leq \theta_{\text{max}}$.
	Furthermore, let $\mu$ be bounded by $\mu < \frac{\chi}{2 \theta_{\text{max}}}$. 
	Then, there are two possible cases.
	\begin{itemize}
		\item Case 1: $y > 0$ and $y + \mu \theta > 0$: \newline
		Since $y + \mu \theta > 0 \Leftrightarrow y > - \mu \theta > - \mu \theta_{\text{max}} > - \frac{\chi}{2 \theta_{\text{max}}} \theta_{\text{max}} = \frac \chi 2$, this case means, that $y \in \tilde{\Y}$.
		Here, we obtain
		$$ \D_\theta V(y) = \liminf_{\mu \rightarrow 0} \frac{y + \mu \theta - y}{\mu} = \theta. $$
		Furthermore, \eqref{eqn:loc-hom-cond} holds, since
		$$ \abs{ \frac{y + \mu' \theta - y}{\mu'} - \D_\theta V(y)} = 0 \leq \nu.$$
		Thus, \eqref{eqn:loc-hom-cond} holds for all $y > \nicefrac \chi 2$.
		
		\item Case 2: $y > 0$ and $y + \mu \theta \leq 0$: \newline
		In this case, $y \in (0,\nicefrac \chi 2]$ and \eqref{eqn:loc-hom-cond} does not hold, since $\D_\theta V(\tilde y) = - \theta + \liminf_{\mu \rightarrow 0} - 2 \nicefrac y \mu$ can not be computed, but based on \eqref{eqn:neighb-pt-cond}, a point $\tilde y \in \tilde{\Y}$ can be chosen.
		Then, this point satisfies \eqref{eqn:loc-hom-cond}, since $\tilde y > \nicefrac \chi 2$ is just case 1.		
	\end{itemize}	
\end{exmp}

\subsection{Proof of Lemma \ref{lem:1-eps-minimizer-vicinity}}

\begin{proof}
	Define $R_1 := (2 \bar V) ^{\nicefrac 1 2} \alpha$.
	Then,
	\begin{align*}
		\inf_{\norm{x - y} \le R_1} \left\{ V(y) + \frac{1}{2 \alpha^2} \norm{y - x}^2 \right\} \leq V(x) \leq \bar V
	\end{align*}
	holds for all $x \in \ball_R$.	
	Furthermore, for any $R_2 > R_1$,
	\begin{align*}
		\inf_{R_1 \le \norm{x - y} \le R_2} \left\{ V(y) + \frac{1}{2 \alpha^2} \norm{y - x}^2 \right\} \geq \frac{1}{2 \alpha^2} R_1^2 \geq \bar V
	\end{align*}
	holds as well.
	Therefore, 
	\begin{align*}
		\inf_{y \in \R^n} \left\{ V(y) + \frac{\norm{y - x}^2}{2 \alpha^2} \right\} = \inf_{\norm{x - y} \le R_1} \left\{ V(y) + \frac{\norm{y - x}^2}{2 \alpha^2} \right\}.
	\end{align*}
\end{proof}

\subsection{Proof of Lemma \ref{lem:2-accuracy-InfC}}

\begin{proof}
	The first inequality follows directly from the definition of the InfC according to \eqref {eqn:InfC}.
	Lemma \ref{lem:1-eps-minimizer-vicinity} implies $\norm{y_\alpha^\eps (x) - x} \leq (2 \bar V)^{\nicefrac 1 2}$, since $\alpha < 1$.
	Choose $\eps_1$ such that $(2 \bar V)^{\nicefrac 1 2} \leq \frac{\eps_1}{2 L_V}$.
	Then, by Lipschitzness of $V$, $\abs{V(x) - V(y_\alpha^\eps(x))} \leq L_V \norm{x - y_\alpha^\eps(x)} \leq \nicefrac{\eps_1}{2}$ follows, and also $V(x) - V(y_\alpha^\eps(x)) \leq \nicefrac{\eps_1}{2} \Leftrightarrow V(x) \leq V(y_\alpha^\eps(x)) + \nicefrac{\eps_1}{2}$.
	Furthermore, \eqref{eqn:InfC-approx-optim} yields $V(y_\alpha^\eps(x)) \leq V_\alpha(x) + \eps \leq V_\alpha(x) + \nicefrac{\eps_1}{2}$.
	Combining these two inequalities yields the desired result.
\end{proof}

\subsection{Proof of Theorem \ref{thm:pract-stab}}

\begin{proof}
	The proof is split into four parts.
	Preliminary settings are made in the first part. 
	In the second one, a relaxed decay condition of the CLF and InfC is presented.
	The actual decay is demonstrated in the third part and in the last part, the parameters for the decay are determined.	 

	\subsection*{Part 1: Preliminaries} \label{sub:proof-part-1}
	
	Let $\ball_r$ be the target and $\ball_R$ the starting ball for $x$, respectively. 
	
	Construct two non-decreasing functions $\varrho_V $ and $\lambda_V$ with the properties 
	\begin{equation} 
		\label{eqn:Func-rho-V}
		\forall x \in \R^n, r, v > 0: V(x) \le \varrho_V(r) \implies \norm x \le r
	\end{equation}
	and
	\begin{equation} 
		\label{eqn:Func-lambda-V}
		V(x) \ge v \implies \norm x \geq \lambda_V(v).
	\end{equation}
	
	By Lemma 4.3 in \cite{Khalil2002-nonlin-sys}, there exist two class $\mathcal K_\infty$ functions $\alpha_1$ and $\alpha_2$ \sut $V(x)$ can be bounded via $\alpha_1(\norm x) \le V(x) \le \alpha_2(\norm x), \spc \forall x \in \R^n$.
	Taking $\varrho_V(r)$ as $\alpha_1(r)$ and $\lambda_V(r)$ as $\alpha_2^{-1}(r)$ yield the above properties. 
	Due to Lemma \ref{lem:2-accuracy-InfC}, \eqref{eqn:V-V-alpha-approx} holds for any $\eps_1 \in \R$.
	It follows that 
	\begin{equation}
		V_\alpha(x) \le \varrho_V(r) - \eps_1 \implies V(x) \le \varrho_V(r) \implies \norm x \le r
	\end{equation}
	and
	\begin{equation}
		V_\alpha(x) \ge v \implies V(x) \ge v \implies \norm x \ge \lambda_V(v).
	\end{equation}
	Let $q$ and $e$ be bounded from above by $\bar q \le \frac{r}{8}$ and, respectively, $\bar e \le \frac{r}{8}$ for all $t \ge 0$ according to Assumption \ref{asm:sys-props}.

	Define $\hat R := R + \bar e + \bar q$, which is given as the radius of the starting ball for $\hat x$ and set $\hat V := \sup_{\norm x \le \hat R} V(x)$.
	Choose $\hat R^*$ and define $\Theta$ such that $\hat V \le \Theta := \varrho_V(\hat R^*)$ holds.	
	If $V(\hat x) \le \varrho_V(\hat R^*)$, then $\norm{\hat x} \le \hat R^*$ and, furthermore, $\norm x \le R^* := \hat R^* + \bar e$.
	Thus, $\hat R^*$ yields an overshoot bound for the measured state $\hat x$ and $R^*$ is given as an overshoot bound for the real state $x$.
	Define $\hat V^* := \sup_{\norm x \le \hat R^*} V(x)$. \newline
	Let $\hat r:= r - \bar e - \bar q$ be the radius of the target ball for $\hat x$ and define $\hat v := \varrho_V(\hat r)$.
	Then, $V(\hat x) \le \varrho_V(\hat r)$ implies $\norm{\hat x} \le \hat r$ and $\norm x \le r$.
	Set $\hat r^* := \lambda_V(\nicefrac{\hat v}{4})$, which is denoted as the radius of a ball, never be entered by $\hat x(t)$.
	
	It follows, that $V(\hat x) \ge \nicefrac{\hat v}{4}$ implies $\norm{\hat x} \ge \hat r^*$ and $\norm x \ge r^* := \hat r^* - \bar e$.	
	
	Let $\U^* \subseteq \U$ be the compact set corresponding to $\ball_{\hat R^* + \sqrt{2 \hat V^*}}$ in \eqref{eqn:decay}. 
	Then,
	\begin{equation} \label{eqn:w(x)}
		\forall x \in \ball_{\hat R^* + \sqrt{2 \hat V^*}}: \inf_{ \theta \in \co(f(x, \U^*)) } \D_\theta V(x) \le -w(x).
	\end{equation}
	Let $L_f$ be the Lipschitz constant of $f$ on $\ball_{\hat R^* + \sqrt{2 \hat V^*}}$.
	Finally, set 
	\begin{align} 
		\label{eqn:bar-f-bar-w}
		& \bar f := \sup_{ \substack { x \in \ball_{\hat R^* + \sqrt{2 \hat V^*}} \\   u \in \U^*} } \norm{f(x, u)}, \spc 
		& \bar w := \inf_{ \frac{\hat r^*}{2} \le \norm x  \le \hat R^* + \sqrt{2 \hat V^*}} w(x),
	\end{align} 
	and consider the Lipschitz condition for the CLF with $\abs{V(y) - V(x)} \le L_V \norm{y - x}, \spc \forall x, y \in \ball_{\hat R^*+\sqrt{2 \hat V^*}}$.
	
	\begin{figure}
		\centering
		\includegraphics[width = 0.5\textwidth]{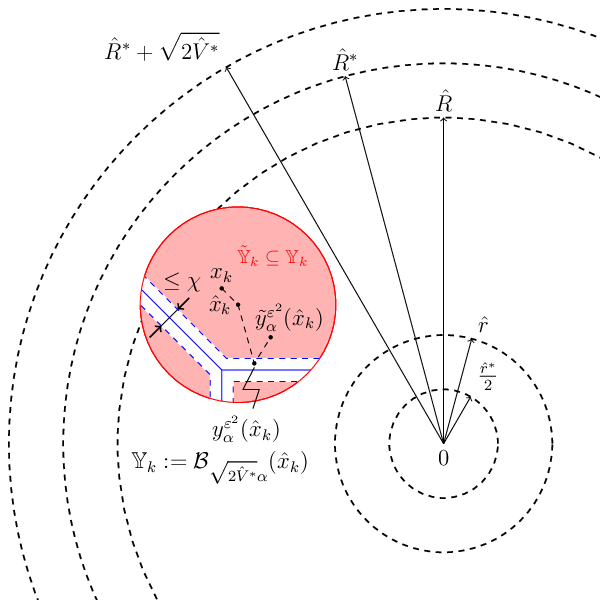}
		\small
		\begin{tabular}{llll}
			$x_k$ & True state & $\hat x_k$ & Measured state \\
			$y_\alpha^{\eps^2}(\hat x_k)$ & Approximative & $\tilde y _\alpha^{\eps^2}(\hat x_k)$ & A point \\
			  & minimizer of \eqref{eqn:InfC} & & satisfying \eqref{eqn:neighb-pt-cond} \\
			$\ball_{R^*}$ & Overshoot & $\ball_{\hat R^*}$ & Overshoot \\
			 & bound ($x$)  & & bound ($\hat x$) \\
			$\ball_{\hat R} $ & Starting ball ($\hat x$) & $\ball_R$ & Starting ball ($x$) \\
			$\ball_r$ & Target ball ($x$) & $\ball_{\hat r}$ & Target ball ($\hat x$) \\
			$\ball_{\nicefrac{\hat r^*}{2}}$ & Core ball ($\hat x$) & $\ball_{\nicefrac{r^*}{2}}$ & Core ball ($x$)
		\end{tabular} 
		\caption{A schematic picture of the geometric setting of the proof.}
		\label{fig:balls}
	\end{figure}
	
	\subsection*{Part 2: Establishing decay rate} \label{sub:proof-part-2}
	
	Consider $x \in \ball_{\hat R^* + \sqrt{2 \hat V^*}}$.
	Let $y_\alpha^{\eps^2}(\hat x)$ be an approximate minimizer of $V_\alpha(x)$ satisfying \eqref{eqn:InfC-approx-optim} and define the corresponding proximal $\eps_x^2$-subgradient $\zeta_\alpha^{\eps^2}(\hat x)$ as $\zeta_\alpha^{\eps^2}(\hat x) := \frac{ \hat x - y_\alpha^{\eps^2}(\hat x) }{\alpha^2}$.
	The minimizer $y_\alpha^{\eps^2}(\hat x)$ must not necessarily satisfy \eqref{eqn:loc-hom-cond}, but based on \eqref{eqn:neighb-pt-cond}, a point $\tilde y_\alpha^{\eps^2}(\hat x)$ in a ball of radius $\chi$ centered at the minimizer can be found \sut \eqref{eqn:loc-hom-cond} holds. 
	It means, that $\tilde y_\alpha^{\eps^2}(\hat x) \in \ball_\chi (y_\alpha^{\eps^2}(\hat x))$ \ie that this point is within a $\chi$-ball of the respective approximate minimizer.
	It is used to define $\tilde\zeta_\alpha^{\eps^2}(\hat x) := \frac{ \hat x - \tilde y_\alpha^{\eps^2}(\hat x) }{\alpha^2}$. 
	In the following, a decay condition will be established for the scalar product $\scal{ \tilde\zeta_\alpha^{\eps^2}(\hat x), f(\hat x, \tilde\kappa_{\hat x}^{\eta}) }$, where $\tilde\kappa^\eta_{\hat x} \in \U^*$ is given as a control law satisfying \eqref{eqn:ctrl-approx-optim} for a given $\eta_{\hat x}$.
	
	The parameters $\eps_{\hat x}$ and $\eta_{\hat x}$ will be determined later. 
	
	With the help of the Lipschitz constant $L_f$, equations \eqref{eqn:Lip-cond} and \eqref{eqn:ctrl-approx-optim}, the following inequality holds for any $\hat x \in \ball_{\hat R^*} \sm \ball_{\hat r^*}:$ 
		\begin{equation}
			\label{eqn:Scal-zeta-f}
		\begin{split}
				&\scal{ \tilde\zeta_\alpha^{\eps^2}(\hat x), f(\hat x, \tilde\kappa_{\hat x}^\eta) } \\
				&  \ \ \ \ \ \ \ \ \ \ \ \ \ = \scal{ \tilde\zeta_\alpha^{\eps^2}(\hat x), f(\tilde y_\alpha^{\eps^2}(\hat x), \tilde\kappa_{\hat x}^\eta) } \\
				&  \ \ \ \ \ \ \ \ \ \ \ \ \  \ \ \ + \scal{ \tilde\zeta_\alpha^{\eps^2}(\hat x), f(\hat x, \tilde\kappa_{\hat x}^\eta) - f(\tilde y_\alpha^{\eps^2}(\hat x), \tilde\kappa_{\hat x}^{\eta}) } \\
				&  \ \ \ \ \ \ \ \ \ \ \ \ \  \le \inf_{ u \in \U^* } \scal{ \tilde\zeta_\alpha^{\eps^2}(\hat x), f(\tilde y_\alpha^{\eps^2}(\hat x), u) } + \eta_{\hat x} \\
				&  \ \ \ \ \ \ \ \ \ \ \ \ \  \ \ \ + \norm{ \tilde\zeta_\alpha^{\eps^2}(\hat x) } L_f \norm{\hat x - \tilde y_\alpha^{\eps^2}(\hat x)}.
			\end{split}
		\end{equation}
		Notice that $y_\alpha^{\eps^2}(\hat x)$ is an $\eps_{\hat x}^2$-minimizer for the inf-convolution \eqref{eqn:InfC}.
		The control actions are determined in an approximate format characterized by $\eta_{\hat x}$. 
		For now, using the relations \eqref{eqn:InfC-approx-optim}, \eqref{eqn:subgrad-zeta-approx-optim}, \eqref{eqn:neighb-pt-cond}, and the definition of $\tilde \zeta_\alpha^{\eps^2}(\hat x)$, an upper bound for $\norm{ \tilde\zeta_\alpha^{\eps^2}(\hat x) } \norm{ \tilde y_\alpha^{\eps^2}(\hat x) - \hat x }$ in \eqref{eqn:Scal-zeta-f} can be determined by
		\begin{equation} 
			\label{eqn:norm-zeta-yx}
			\begin{split}
			&\norm{ \tilde\zeta_\alpha^{\eps^2}(\hat x) } \norm{ \tilde y_\alpha^{\eps^2}(\hat x) - \hat x } = \frac{1}{\alpha^2} \norm{ \tilde y_\alpha^{\eps^2}(\hat x) - \hat x }^2 \\
				&\le \frac{1}{\alpha^2} \left( \norm{ \tilde y_\alpha^{\eps^2}(\hat x) - y_\alpha^{\eps^2}(\hat x)} + \norm{y_\alpha^{\eps^2}(\hat x) - \hat x} \right)^2 \\
				&\le \frac{2}{\alpha^2} \left( \norm{ \tilde y_\alpha^{\eps^2}(\hat x) - y_\alpha^{\eps^2}(\hat x)}^2 + \norm{y_\alpha^{\eps^2}(\hat x) - \hat x}^2 \right) \\
				&\le \frac{2}{\alpha^2} (\chi^2 + 2 \alpha^2 ( V(\hat x) - V(y_\alpha^{\eps^2}(\hat x)) + \eps_{\hat x}^2)). \\
			\end{split}
		\end{equation}
		The second inequality in \eqref{eqn:norm-zeta-yx} result from the fact, that for any $a,b > 0$, it holds that $(a + b)^2 = a^2 + b^2 + 2ab \le 2a^2 + 2b^2$, since $0 \le (a - b)^2 = a^2+ b^2 - 2ab \Leftrightarrow 2ab \le a^2 + b^2$.
		Combining \eqref{eqn:Scal-zeta-f} and \eqref{eqn:norm-zeta-yx} and choosing $\alpha$ such that $\sqrt{2 \hat V^*} \alpha \le \left\{ \frac{\hat r^*}{2}, \frac{\eps_1}{L_V} \right\}$ holds, the following equation can be obtained, which holds for all $\hat x \in \ball_{\hat R^*} \sm \ball_{\hat r^*}$:
			\begin{equation} 	
				\label{eqn:Scal-zeta-f-final}
				\begin{split}		
					&\scal{ \tilde\zeta_\alpha^{\eps^2}(\hat x), f(\hat x, \tilde\kappa_{\hat x}^{\eta}) } \\
					&\le \inf_{ u \in \U^* } \scal{ \tilde\zeta_\alpha^{\eps^2}(\hat x), f(\tilde y_\alpha^{\eps^2}(\hat x), u) } + \eta_{\hat x}  \\
					& \ \ \ + L_f \left( \frac{2}{\alpha^2} (\chi^2 + 2 \alpha^2 ( V(\hat x) - V(y_\alpha^{\eps^2}(\hat x)) + \eps_{\hat x}^2)) \right) \\
					&\le \inf_{ u \in \U^* } \scal{ \tilde\zeta_\alpha^{\eps^2}(\hat x), f(\tilde y_\alpha^{\eps^2}(\hat x), u) } + \eta_{\hat x}  \\
					& \ \ \ + L_f \left( \frac{2}{\alpha^2} (\chi^2 + 2 \alpha^2 ( \eps_1 + \eps_{\hat x}^2)) \right).
 				\end{split}
			\end{equation}
		
	Furthermore, observe that $\norm{ y_\alpha^{\eps^2} (\hat x) } \in [\nicefrac{\hat r^*}{2}, \hat R^* + \sqrt{2 \hat V^*}]$, since $\hat x \in \ball_{\hat R^*} \sm \ball_{\hat r^*}$.
	
	\subsection*{Part 3: Deriving decay along system trajectories} \label{sub:proof-part-3}
	
	Consider an arbitrary $\hat x \in \X$, where $\X \subset \R^n$ is compact.
	For its subgradient $\zeta_\alpha^{\eps^2}(\hat x)$, the following condition holds for any $\hat x \in \X$ and $h \in \R, \theta\in \R^n$:
			\begin{equation} 
				\label{eqn:Taylor-expansion}
					V_\alpha(\hat x + h \theta)\le V_\alpha(\hat x) + h \scal{ \zeta_\alpha^{\eps^2}(\hat x), \theta } + \frac{h^2 \norm\theta^2}{2 \alpha^2} + \eps_{\hat x}^2.
			\end{equation}	
	Note that \eqref{eqn:Taylor-expansion} does not hold for $\tilde\zeta_\alpha^{\eps^2}(\hat x)$ instead of $\zeta_\alpha^{\eps^2}(\hat x)$, since it is not even an approximative proximal subgradient. 		
		Therefore, observe that for all $\hat x \in \ball_{\hat R^* + \sqrt{2 \hat V^*}}$:
		\begin{equation}
			\label{eqn:subgradient-zeta-tilde}
			\begin{split}
				\zeta_\alpha^{\eps^2}(\hat x) &= \frac{\hat x - y_\alpha^{\eps^2}(\hat x)}{\alpha^2} 
				=\frac{\hat x - \tilde y_\alpha^{\eps^2}(\hat x) + \tilde y_\alpha^{\eps^2}(\hat x) - y_\alpha^{\eps^2}(\hat x)}{\alpha^2} \\
				&= \tilde\zeta_\alpha^{\eps^2}(\hat x) + \frac{ \tilde y_\alpha^{\eps^2}(\hat x) - y_\alpha^{\eps^2}(\hat x) }{\alpha^2}
			\end{split}
		\end{equation}
		holds. 
		Furthermore, based on \eqref{eqn:neighb-pt-cond}, $\norm{ \tilde y_\alpha^{\eps^2}(\hat x) - y_\alpha^{\eps^2}(\hat x) } \leq \chi$ holds as well.
		Consider Taylor expansion \eqref{eqn:Taylor-expansion} and \eqref{eqn:subgradient-zeta-tilde}.
		Then, the following inequalities hold:
		\begin{equation}
			\label{eqn:Taylor-exp}
		\begin{split}
			&V_\alpha(\hat x + h \theta) \\
			&\le V_\alpha(\hat x) + h \scal{ \zeta_\alpha^{\eps^2}(\hat x), \theta } + \frac{h^2 \norm\theta^2}{2 \alpha^2} + \eps_{\hat x}^2 \\
				&\le V_\alpha(\hat x) + h \scal{ \tilde\zeta_\alpha^{\eps^2}(\hat x), \theta } + \frac{h^2 \norm\theta^2}{2 \alpha^2} + \eps_{\hat x}^2 + h \frac{\chi}{\alpha^2} \norm\theta.
			\end{split}
		\end{equation}
	Assume now that the trajectory of \eqref{eqn:sys-SH} exists locally on the sampling period $[k \delta, (k + 1) \delta]$ and that $V_\alpha(\hat x_k)\le \hat V$ holds.
	To see that it exists on the entire sampling period, observe that, based on Lemma \ref{lem:2-accuracy-InfC} with
	\begin{equation}
		V_\alpha(\hat x) \le V(\hat x) \le V_\alpha(\hat x) + \eps_1, \forall \hat x \in \ball_{\hat R^*},
	\end{equation}
	the following inequalities hold for $t \in [k \delta, (k + 1) \delta]$:
	\begin{equation}
		\label{eqn:V-alpha-bounds}
		V(\hat x (t)) \le V_\alpha(\hat x (t)) + \eps_1 \le \hat V + \eps_1.
	\end{equation}
	Inequality \eqref{eqn:V-alpha-bounds} is used to show that the trajectory $\hat x (t)$ exists on the entire sampling period and it can be also used to find a bound for $\eps_1$ to satisfy $V(\hat x (t)) \le \Theta$ which implies $\norm{\hat x (t)} \le \hat R^*$ and that means, that the overshoot is bounded, $\hat x(t) \in \ball_{\hat R^*}$ and $x(t) \in \ball_{R^*}$, for all $t \ge 0$. 
	It is shown in the following steps, that $V_\alpha(\hat x_k)$ can only decrease to a prescribed limit sample-wise \ie $V_\alpha(\hat x_{k+1}) \le V_\alpha(\hat x_k)$ for $k \in \N$ until $V_\alpha(\hat x_k) \le \hat v$.
	This ensures the boundedness of the trajectory at each sampling period.
	
	Now, consider the following cases. 
	
	\textit{Case 1:} $V_\alpha(\hat x_k) \ge \frac{\hat v}{2}$
	\ (Outside the core ball) \newline
	The trajectory $\hat x(t)$ can be expressed as
	\begin{equation} 
		\label{eqn:X-t-hat}
		\begin{split}
			\hat x (t) &= \hat x_k + \int_{k \delta}^t f(\hat x (\tau), \tilde\kappa_{\hat x_k}^\eta) + q(\tau) \spc \mathrm d \tau \\
			&= \hat x_k + \delta \underbrace{\frac{1}{\delta} \left( \int_{k \delta}^t f(\hat x (\tau), \tilde\kappa_{\hat x_k}^{\eta}) + q(\tau) \spc \mathrm d \tau \right) }_{=:F_k}.
		\end{split}
	\end{equation}	
	Furthermore, the following inequality can be obtained using \eqref{eqn:Taylor-exp}:
	\begin{equation}
		\label{eqn:Difference-LF-value-states}
		\begin{split}
			&V_\alpha(\hat x (t)) - V_\alpha(\hat x_k) = V_\alpha(\hat x_k + \delta F_k) - V_\alpha(\hat x_k) \\
			&\le \delta \scal{ \tilde\zeta_\alpha^{\eps^2}(\hat x_k), F_k } + \frac{\delta^2 \norm{F_k}^2}{2 \alpha^2} + \eps_{\hat x_k}^2 + \delta \frac{\chi}{\alpha^2} \norm{F_k}
		\end{split}
	\end{equation}
	for all $t \in [k \delta, (k + 1) \delta]$ with $\Delta t := t - k \delta$.
	Since $F_k$ can be bounded as $\norm{F_k} \le \frac{1}{\delta} \Delta t (\bar f + \bar q)$, it  can be re-expressed as
	\begin{equation}	
		\label{eqn:F-k}
		\begin{split}
			F_k &= \frac{\Delta t}{\delta} f(\hat x_k, \tilde\kappa_{\hat x_k}^\eta) + \frac{1}{\delta} \int _{k \delta}^t q(\tau) \spc \mathrm d \tau  \\
			& \ \ \ + \underbrace{ \frac{1}{\delta} \int_{k \delta}^t f(\hat x(\tau), \tilde\kappa_{\hat x_k}^\eta) - f(\hat x_k, \tilde\kappa_{\hat x_k}^\eta) \spc \mathrm d \tau }_{=:A}
		\end{split}
	\end{equation}
	and $\norm A \le \frac{1}{\delta} \Delta t^2 L_f \bar f$, where $\bar q$ is bounded later.
	Under Lemma \ref{lem:1-eps-minimizer-vicinity}, equation \eqref{eqn:F-k}, inequality \eqref{eqn:Scal-zeta-f-final} and the definition of $\tilde \zeta_\alpha^{\eps^2}(\hat x)$, it follows that
	\begin{equation} 
		\label{eqn:Scal-zeta-Fk-inequ}
		\begin{split}
			&\scal{ \tilde\zeta_\alpha^{\eps^2}(\hat x_k), F_k } \\
			&= \scal{ \tilde\zeta_\alpha^{\eps^2}(\hat x_k), \frac{\Delta t}{\delta} f(\hat x_k, \tilde\kappa_{\hat x_k}^\eta) } \\
			& \ \ \ + \scal{ \tilde\zeta_\alpha^{\eps^2}(\hat x_k), A + \frac{1}{\delta} \int_{k \delta}^t q(\tau) \ \mathrm d \tau} \\
			&\le \frac{\Delta t}{\delta} \scal{ \tilde\zeta_\alpha^{\eps^2}(\hat x_k), f(\hat x_k, \tilde\kappa_{\hat x_k}^\eta) } \\
			& \ \ \ + \norm{ \tilde\zeta_\alpha^{\eps^2}(\hat x_k) } \left(\frac{\Delta t^2}{\delta} L_f \bar f + \frac{\Delta t}{\delta} \bar q \right) \\
			&\le \frac{\Delta t}{\delta} \left( \inf_{ u \in \U^* } \scal{ \tilde\zeta_\alpha^{\eps^2}(\hat x_k), f(\tilde y_\alpha^{\eps^2}(\hat x_k), u) } + \eta_{\hat x_k} \right. \\
			& \ \ \ + \left. L_f \left( \frac{2}{\alpha^2} \chi^2 + 4 ( \eps_1 + \eps_{\hat x_k}^2) \right) \right) \\
			& \ \ \ + \left( \frac{\sqrt{2 \hat V^*}}{\alpha} + \frac{\chi}{\alpha^2} \right) \left( \frac{\Delta t^2}{\delta} L_f \bar f + \frac{\Delta t}{\delta} \bar q \right).
		\end{split}
	\end{equation}
	For $t = (k + 1) \delta$ the following inequality can be obtained with \eqref{eqn:Difference-LF-value-states}:
	\begin{equation}
		\label{eqn:diff-V-alpha-time-step}
		\begin{split}
			&V_\alpha(\hat x_{k+1}) - V_\alpha(\hat x_k) \\
			&\le \delta \left[ \inf_{ u \in \U^* } \scal{ \tilde\zeta_\alpha^{\eps^2}(\hat x_k), f(\tilde y_\alpha^{\eps^2}(\hat x_k), u) } + \eta_{\hat x_k} \right. \\
			& \ \ \ + \left. (\delta L_f \bar f + \bar q) \frac{\chi}{\alpha^2} + L_f \left( \frac{2}{\alpha^2} \chi^2 + 4 ( \eps_1 + \eps_{\hat x_k}^2) \right) \right. \\
			& \ \ \ + \left. (\delta L_f \bar f + \bar q) \frac{\sqrt{2 \hat V^*}}{\alpha} \right]  \\
			& \ \ \ + \frac{\delta^2 (\bar f + \bar q)^2}{2 \alpha^2} + \eps_{\hat x_k}^2 + \frac{\chi}{\alpha^2} \delta (\bar f + \bar q).
		\end{split}
	\end{equation}
	
	\textit{Case 2:} $V_\alpha(\hat x_k) \le \frac{3}{4}\hat v$
	\ (Inside the target ball) \newline
	If the sample period size $\delta$ satisfies $\delta \bar f \le \frac{\eps_2}{L_V}$ for some $\eps_2 > 0$, then $V_\alpha(\hat x (t)) \le V_\alpha(\hat x_k) + \eps_2$.
	Choosing $\eps_2 \le \frac{\hat v}{8}$ guarantees that $V_\alpha(\hat x (t)) \le \frac{ 7 \hat v}{8}$, and $\eps_1$ satisfying $V(\hat x (t)) \le \hat v$ ensures $\norm{\hat x (t)} \le \hat r$ and $\norm{x(t)} \le r$ for all $t \ge 0$.
	
	\subsection*{Part 4: Determining parameters for decay} \label{sub:proof-part-4}
	
	Some of the parameters \eg $\eps_1$ and $\eps_2$, were already determined in the previous parts. 
	In the following, the different summands of \eqref{eqn:diff-V-alpha-time-step} are bounded.
	With \eqref{eqn:diff-V-alpha-time-step} and $\delta < 1$, $\eps_1$ needs to satisfy
	\begin{equation} 
		\label{eqn:Bounds-1}
		4 L_f \varepsilon_1 \le \frac{\bar w}{36}.
	\end{equation}		
	Note that these bounds influence also $\eps_{\hat x_k}$ indirectly.
	Fix $\alpha$ and set the following bounds
	\begin{equation}	
		\label{eqn:Bounds-2}
		\eta_{\hat x_k} \le \frac{\bar w}{36}, \spc
		\delta\frac{\bar w}{2} \le \frac{\hat v}{4}, \spc
		\frac{\delta (\bar f + \bar q)^2}{2 \alpha^2} \le \frac{\bar w}{36}.
	\end{equation}
	Force $\delta$ to additionally satisfy $(\delta L_f \bar f + \bar q) \frac{\sqrt{2 \hat V^*}}{\alpha} \le \frac{\bar w}{36}$.
	From now on, $\delta$ is considered fixed and $\eps_{\hat x_k}^2$ is constrained by 
	\begin{equation}
		\label{eqn:Bounds-3}
		\eps_{\hat x_k}^2 \le \delta \frac{\bar w}{36}, \spc
		4 L_f \eps_{\hat x_k}^2 \le \frac{\bar w}{36}.
	\end{equation} 
	Furthermore, the following inequalities should hold:
	\begin{equation}
		\label{eqn:Bounds-4}
		\frac{2}{\alpha^2} L_f \chi^2 \le \frac{\bar w}{36}, \spc
		(\delta L_f \bar f + \bar q) \frac{\chi}{\alpha^2} \le \frac{\bar w}{36}, \spc
		\frac{\chi}{\alpha^2} (\bar f + \bar q) \le \frac{\bar w}{36}.
	\end{equation}
	Now, bounds on the optimization precision $\eps_{\hat x_k}^2$ are derived to achieve
	\begin{equation}
		\inf_{ u \in \U^* } \scal{ \tilde\zeta_\alpha^{\eps^2}(\hat x_k), f(\tilde y_\alpha^{\eps^2}(\hat x_k), u) } \le - \frac{3 \bar w}{4}.
	\end{equation}
	To this end, observe that based on \eqref{eqn:Taylor-exp}, for all $z \in \R^n$,
		\begin{equation}
			\begin{split}
				V(z) \ge &V(\tilde y_\alpha^{\eps^2}(\hat x_k)) + \scal{ \tilde\zeta_\alpha^{\eps^2}(\hat x_k), z - \tilde y_\alpha^{\eps^2}(\hat x_k) } \\
				&- \frac{1}{2 \alpha^2} \norm{ z - \tilde y_\alpha^{\eps^2}(\hat x_k) }^2 - \eps_{\hat x_k}^2  \\
				&- \underbrace{\scal{\frac{\tilde y_\alpha^{\eps^2} (\hat x_k) - y_\alpha^{\eps^2} (\hat x_k)}{\alpha^2} , z - \tilde y_\alpha^{\eps^2} (\hat x_k)}}_{\leq \frac{\chi}{\alpha^2} \norm{z - \tilde y_\alpha^{\eps^2} (\hat x_k)}}
			\end{split}
		\end{equation}
		holds and also, for any $\theta \in \R^n$,
		\begin{equation} 
			\begin{split}
				V( \tilde y_\alpha^{\eps^2}(\hat x_k) + \eps_{\hat x_k}\theta ) \ge &V( \tilde y_\alpha^{\eps^2}(\hat x_k) ) + \eps_{\hat x_k} \scal{ \tilde\zeta_\alpha^{\eps^2}(\hat x_k), \theta } \\
				&- \frac{1}{2 \alpha^2} \eps_{\hat x_k}^2 \norm\theta^2 - \eps_{\hat x_k}^2 - \frac{\chi}{\alpha} \eps_{\hat x_k} \norm \theta. 
			\end{split}
		\end{equation}
		This inequality yields the following bound:
		\begin{equation} 
			\label{eqn:scal-zeta-theta+q}
			\begin{split}
				\scal{ \tilde\zeta_\alpha^{\eps^2}(\hat x_k), \theta } \le &\frac{ V( \tilde y_\alpha^{\eps^2}(\hat x_k) + \eps_{\hat x_k}\theta ) - V( \tilde y_\alpha^{\eps^2}(\hat x_k) ) }{\eps_{\hat x_k}} \\
				&+ \frac{1}{2 \alpha^2} \eps_{\hat x_k} \norm\theta^2 + \frac{\chi}{\alpha^2} \norm \theta + \eps_{\hat x_k}. 
			\end{split}
		\end{equation}
		Using Lemma \ref{lem:1-eps-minimizer-vicinity} and Assumption \ref{asm:separable-edges} (which ensures $\tilde y_\alpha^{\eps^2}(\hat x_k) \in \tilde{\Y}_k \subseteq \Y_k := \ball_{\sqrt{2 \hat V^*} \alpha} (\hat x_k)$) enables, for $\eps_{\hat x_k}^2 < \mu$, the condition
		\begin{equation} 
			\label{eqn:Homogenity-w/5}
			\begin{split}
				&\abs{ \frac{ V( \tilde y_\alpha^{\eps^2}(\hat x_k) + \eps_{\hat x_k}^2 \theta ) - V( \tilde y_\alpha^{\eps^2}(\hat x_k) ) }{\eps_{\hat x_k}^2} - \D_\theta V( \tilde y_\alpha^{\eps^2}(\hat x_k) ) } \\
				&\le \frac{\bar w}{5}, \spc \forall \theta \in \co(f( \tilde y_\alpha^{\eps^2}(\hat x_k), \U^* ) ).
			\end{split}
		\end{equation}
		With \eqref{eqn:scal-zeta-theta+q} and \eqref{eqn:Homogenity-w/5}, it yields 
		\begin{equation} 
			\label{eqn:Scal-zeta-f-inequ}
			\begin{split}
				&\scal{ \tilde\zeta_\alpha^{\eps^2}(\hat x_k), f(\tilde y_\alpha^{\eps^2}(\hat x_k), u) } \le \D_{ f(\tilde y_\alpha^{\eps^2}(\hat x_k), u) } V(\tilde y_\alpha^{\eps^2}(\hat x_k)) + \frac{\bar w}{5} \\
				&+ \frac{\eps_{\hat x_k}}{2 \alpha^2} \norm{ f(\tilde y_\alpha^{\eps^2}(\hat x_k), u) }^2 + \frac{\chi}{\alpha^2} \norm{ f(\tilde y_\alpha^{\eps^2}(\hat x_k), u) } + \eps_{\hat x_k} 
			\end{split} 
		\end{equation}
		for all $u \in \U^*$.
		Consequently, it holds that
		\begin{equation} 
			\label{eqn:Scal-zeta-f-inf}
			\begin{split}
				&\inf_{ \theta \in \co(f( \tilde y_\alpha^{\eps^2}(\hat x_k),\U^* )) } \scal{ \tilde\zeta_\alpha^{\eps^2}(\hat x_k),\theta } \\
				&\le \inf_{ \theta \in \co(f( \tilde y_\alpha^{\eps^2}(\hat x_k), \U^* )) } \D_\theta V( \tilde y_\alpha^{\eps^2}(\hat x_k) ) + \frac{\bar w}{5} \\
				& \ \ \ + \frac{\eps_{\hat x_k}}{2 \alpha^2} \bar f^2 + \frac{\chi}{\alpha^2} \bar f + \eps_{\hat x_k} 
				\le -\frac{4 \bar w}{5} + \frac{\eps_{\hat x_k}}{2 \alpha^2} \bar f^2 + \frac{\chi}{\alpha^2} \bar f + \eps_{\hat x_k}. 
			\end{split}
		\end{equation}
		If $\eps_{\hat x_k}$ is bounded from above via
		\begin{equation} \label{eqn:bound-5}
			\eps_{\hat x_k} \le \frac{ \nicefrac{\bar w \alpha^2}{10} - 2 \chi \bar f}{2 \alpha^2 + \bar f^2},
		\end{equation}		
		the desired result follows:
		\begin{equation}
			\label{eqn:Scal-zeta-f-inequ-final}
			\inf_{ \theta \in \co(f( \tilde y_\alpha^{\eps^2}(\hat x_k), \U^* )) } \scal{ \tilde\zeta_\alpha^{\eps^2}(\hat x_k), \theta } \le -\frac{3}{4}\bar w .
		\end{equation}	
	
	This means that an upper bound for the decay at $\tilde y_\alpha^{\eps^2}(\hat x_k)$ is determined. \newline
	The last step is to show that $V_\alpha(x_{k+1}) \le V_\alpha(x_k)$ for all $k \in \N$.
	For $t = (k + 1) \delta$ an intersample decay rate on $V_\alpha$ for the measured states with bounds \eqref{eqn:diff-V-alpha-time-step}, \eqref{eqn:Bounds-1}-\eqref{eqn:Bounds-4} and \eqref{eqn:Scal-zeta-f-inequ-final} can be established as
	\begin{equation}
		V_\alpha(\hat x_{k+1}) - V_\alpha(\hat x_k) \le \delta \left(-\frac{3}{4} \bar w + \frac{9}{36} \bar w \right) = -\frac{1}{2} \delta \bar w.
	\end{equation}
	Introduce an upper bound for $\bar e$ such that $\bar e < \frac{1}{16} \frac{\bar w}{L_V}$ holds.
		Then, the following inequalities can be obtained
	\begin{equation}
		\begin{split}
			&V_\alpha(x(t)) - V_\alpha(x_k) \\
			&= (V_\alpha(\hat x (t)) - V_\alpha(\hat x_k)) \\
			& \ \ \ + \underbrace{(V_\alpha(\hat x_k) - V_\alpha(x_k))}_{\le L_V \delta \bar e} + \underbrace{(V_\alpha(x(t)) - V_\alpha(\hat x (t)))}_{\le L_V \delta \bar e} \\
			&\le V_\alpha(\hat x (t)) - V_\alpha(\hat x_k) + 2 L_V \delta \bar e, \ \forall t \in [k \delta, (k + 1) \delta].
		\end{split}
	\end{equation}
	The final result for the decay reads as
	\begin{equation}
		V_\alpha(x_{k+1}) - V_\alpha(x_k) \le -\frac{1}{2} \delta \bar w + 2 L_V \delta \bar e < -\frac{3}{8} \delta \bar w. 
	\end{equation}
	This shows that the control action determined in \eqref{eqn:Scal-zeta-f-inequ-final}, computed using $\hat x$, yields a necessary sample-wise decay of $V_\alpha$. 
	
	The reaching time of Case 2 can be determined as $T_\alpha = 2\frac{\hat V^* - \nicefrac{\hat v}{2}}{\delta \bar w}$.
	If Case 2 is reached \ie $V(\hat x_k) \le \frac{3}{4} \hat v$, then two subcases are possible in the following sampling periods. \newline
	Either $\frac{\hat v}{2} \le V_\alpha(\hat x_k) \le \hat v$ (Subcase 2.1) or $V_\alpha(\hat x_k) \le \frac{3 \hat v}{4}$ (Subcase 2.2). 
	If Subcase 2.1 occurs, then $V_\alpha$ can either stay in this subcase during the next sampling period or, based on the decay condition, transition to the other subcase. 
	If the latter subcase occurs, $V_\alpha$ can stay there or move to Case 2.
	Thus, the trajectory $\hat x(t)$ stays in the ball $\ball_{\hat r}$ in any subcase for all the subsequent sampling periods.
	This implies, that $x(t)$ stays in the target ball $\ball_r$ after entering it once.
	This concludes the proof.
\end{proof}

\begin{rem} 
	\label{rem:eps-state-indep}
	In the proof of Theorem \ref{thm:pract-stab}, bounds for optimization precisions $\varepsilon_{\hat x}$ and $\eta_{\hat x}$ are derived depending on the current measured state. 
	To derive uniform bounds would require, in particular, setting $\Y = \ball_R$ and $\F  = f(\Y, \U^*)$.
	Such bounds may be, in general, more conservative than the ones derived in Theorem \ref{thm:pract-stab}.
\end{rem}

The next remark discusses the case when the sampling step size is fixed and the size of the target ball is to be determined.
\begin{rem} 
	\label{rem:bound-optim-acc}
	The current result derives bounds for $\delta, \eps, \eta$ and $\chi$ for a given $r$ and $R$.
	Determining a bound for the radius of the target ball $r$ depending on $R, \delta, \eps$ and $\eta$ would require to consider the bounds \eqref{eqn:Bounds-1}-\eqref{eqn:Bounds-4}, \eqref{eqn:bound-5} as well as the definitions of $\bar w$  and $\bar f$ in \eqref{eqn:bar-f-bar-w}. 
	A particular difficulty is that $\bar w$ is defined on a set which depends on $r$.
	However, if $w$ is independent of $x$ (like in some sliding-mode control setups), the derivation of $r$ may be possible.
\end{rem}

\bibliographystyle{plain}        
\bibliography{bib}

    
\end{document}